# A generalized form of Hamilton's principle

John Hegseth
Department of Physics
University of New Orleans
New Orleans, LA 70148

**Abstract**
Many quantization schemes rely on analogs of classical mechanics where the connections with classical mechanics are indirect. In this work I propose a new and direct connection between classical mechanics and quantum mechanics where the quantum mechanical propagator is derived from a variational principle. I identify this variational principle as a generalized form of Hamilton's principle. This proposed variational principle is unusual because the physical system is allowed to have imperfect information, i.e., there is incomplete knowledge of the physical state. Two distribution functionals over possible generalized momentum paths $\alpha[p(t)]$ and generalized coordinates paths $\beta[q(t)]$ are defined. A generalized action is defined that corresponds to a contraction of $\alpha[p(t)]$, $\beta[q(t)]$, and a matrix of the action evaluated at all possible $p$ and $q$ paths. Hamilton's principle is the extremization of the generalized action over all possible distributions. The normalization of the two distributions allows their values to be negative and they are shown to be the real and imaginary parts of the complex amplitude. The amplitude in the Feynman path integral is shown to be an optimal vector that extremizes the generalized action. This formulation is also shown to be directly applicable to statistical mechanics and I show how irreversible behavior and the micro-canonical ensemble follows immediately.

**Introduction**

Variational principles are appealing for their simplicity and generality [1], [2], [3]. In addition, they also often provide information about a system's stability based on the type of extrema, as in equilibrium statistical mechanics [4]. In classical mechanics there are few general statements about the global character of the extrema, i.e., the type of extrema depends on the details of a specific system [2], [5], e.g., when a system's Hamiltonian $H(p,q,t)$ is a saddle function, the extrema is a global minimum [2], [3], [5]. In this case the action $S[p(t),q(t),t]$ uses the beginning (at time $t_i$) and ending (at time $t_f$) conditions $\delta q(t_i)=0$ and $\delta q(t_f)=0$. As shown in reference [5], finding other general statements again depends on the specific Hamiltonian. Another interesting action $R[p(t),q(t),t]$ was also introduced that uses the beginning and ending conditions $\delta p(t_i)=0$ and $\delta p(t_f)=0$ to derive Hamilton's equations. Both $S$ and $R$, with their respective beginning and ending conditions, result in Hamilton's equations or in a Hamilton-Jacobi equation that determine the motion for perfectly known initial conditions. The same analysis showed that the action $R$ is a global maximum for a saddle function $H(p,q,t)$. In this paper I introduce a variational principle that describes the motion when perfect information about $p$ and $q$ is not available. This is done by generalizing both $S$ and $R$ to include all possible paths and introducing two distributions over both the possible $p(t)$ paths $\alpha[p(t)]$ and the possible $q(t)$ paths $\beta[q(t)]$. First I will show how the paths fan out because of imperfect information. Next I will define the mixed path distribution and show how the mixed path distribution normalizes. I will also show how Hamilton's principle is generalized using these distributions with a generalized action and how the probability amplitude naturally follows from Hamilton's principle. The path integral is next discussed and the probability amplitude is shown to solve this variational principle for the infinite path case. Finally these ideas are shown to directly lead to the micro-canonical distribution in statistical mechanics.

**Imperfect information, distributions, and normalization**

Let experimenters $A$ and $B$ observe a system of particles. I first limit the system to one degree of freedom for simplicity. $A$ is only capable of sampling the momentum path $p(t)$ and $B$ is only able to measure points on the position path $q(t)$. These measurements may be separated by a small sampling time interval $\Delta t$. During $\Delta t$ $A$ observes an initial and final $p$ and knows that $R$ is extremized. During $\Delta t$ $B$ observes an initial and final $q$ and notes that $S$ is extremized. Extremizing either $S$ or $R$ is equivalent to satisfying Hamilton's equations. In addition, there are constraints on the system at a given initial time $t_i$ and a given final time $t_f$. Figure 1 shows an example, during two time intervals, where $q_i$ is given but there are three possible $p_i$ values that $A$ could observe. If either experimenter could simultaneously observe both $p$ and $q$, they would in principle have complete knowledge of the motion and there would not be any indeterminacy. If $A$ and $B$ share information about the initial conditions, i.e., $A$ measures an initial $p = p_i$ and $B$ measures an initial $q = q_i$, then the behavior of the system could be determined by either $A$ or $B$. In the following, a given or fixed point is a constraint on the system and is known in advance by $A$ and $B$, whereas observed or measured points are not known in advance by $A$ or $B$. To represent imperfect information and avoid simultaneous knowledge of $p$ and $q$, I separate $A$'s and $B$'s sampling time by $\Delta t/2$ as shown in Figure 1.

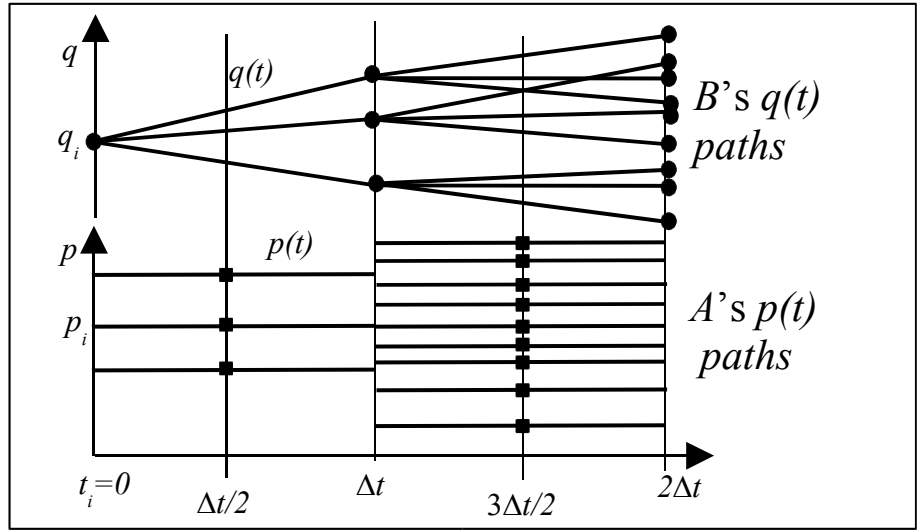

Figure 1. Because there are three possible initial $p$ values, the number of possible paths increases geometrically with time (paths fan out). Each $q(t)$ path that $B$ could observe implies a corresponding $p(t)$ path through Hamilton's equations (and vice versa for $B$).

When there is not complete knowledge or imperfect information, there may be many possible paths. Imperfect information also implies that *A* is completely ignorant of *B*'s measurements and vice versa. Figure 1 shows an example where $q_i$ is given but there are three possible initial values of *p*, i.e., since *B* does not know $p(\Delta t/2)$ he must allow for three possible $q(\Delta t)$. In *A*'s first time interval he finds nine possible values of $p(3\Delta t/2)$ since he must allow for three possible values of $q(\Delta t)$ and three possible initial $p(\Delta t/2)$ before determining the paths. In *A*'s and *B*'s next time interval they find many possible initial *p* and *q* values and evolve the system to $5\Delta t/2$ and $2\Delta t$ for each possible initial condition of the time interval. Clearly there is a very significant increase in the number of possible paths in time (or fan out of paths) as illustrated in Figure 1. If either *A* or *B* fixed an endpoint at a later time, the number of possibilities would be reduced (note that if both of them fixed their endpoints a single path would be determined). If the fixed endpoint $q(t_f)$ occurs at time $T=n\Delta t$ there is significant ignorance of the intermediate positions and momenta, so that many paths are possible. In conclusion, there are many possible *p(t)* and *q(t)* paths, as shown in Figure 1, because the given initial condition is not complete, i.e., *B* has a given initial point while *A* has three possible initial points and neither *A* nor *B* knows the other's measurements.

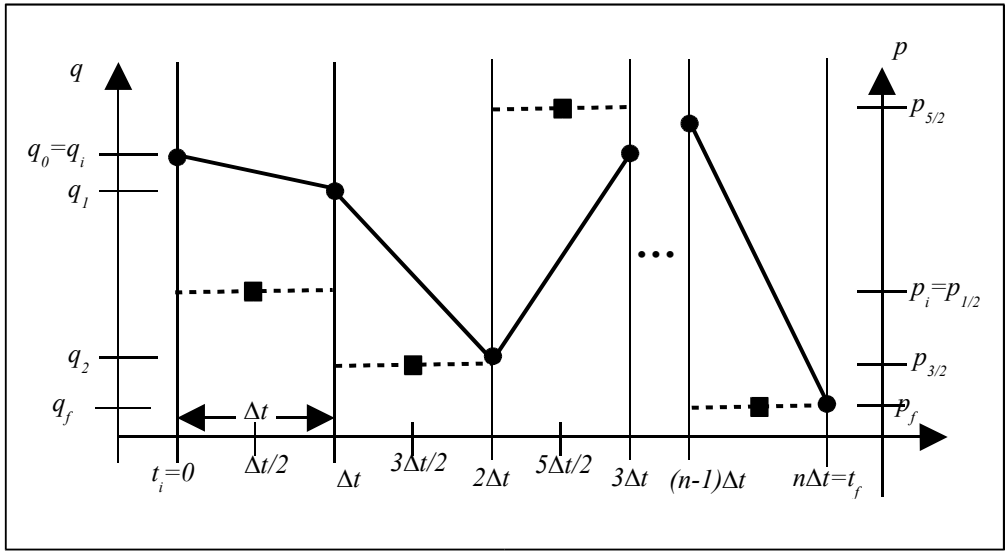

Figure 2. There are two possible ways of representing a path during $T=n\Delta t$, either the paths are manifested in terms of *p(t)* or *q(t)*, as shown above. Both *A and B* could observe many paths between the fixed beginning point and the endpoint. A path is directly observed in one representation while in the other representation the same path is inferred through Hamilton's equations. The distribution functionals $\alpha[p]$ and $\beta[q]$ may be constructed by considering a representative subset of a path as shown above. Each point of the path has a corresponding point value, e.g., $\beta(q_0)$, $\beta(q_1)$, … or $\alpha(p_{1/2})$, $\alpha(p_{3/2})$, …

If *A* were to measure *p(t)*, he may notice through experimental repetition that of the many possible paths some may be more likely than others. Similarly, *B* may observe that some *q(t)* paths may be more likely than others. I next define the distributions for these paths: $\alpha[p(t)]$ for *A* and $\beta[q(t)]$ for *B* between fixed beginning and ending points. I consider the finite time interval $T=n\Delta t$ where the *q(t)* paths fan out at times *0, $\Delta t$, $2\Delta t$, …,*

*(n-1)Δt*. Figure 2 shows one approximate path in terms of both *q(t)* and *p(t)* from this subset of possible paths. It is important to note that for each possible complete path *p(t)* (or *q(t)*) there is a corresponding *q(t)* (or *p(t)*), as illustrated in Figure 2. If *B* knows both *q(t)* and $\dot{q}(t)$, where *q(t)* is approximated by the sequence of points *{q(0), q(Δt),…, q(nΔt)}= {q(iΔt)}$_{i=0,n}$* and $\left\{\frac{q(i\Delta t) - q((i-1)\Delta t)}{\Delta t}\right\}_{i=1,n}$ approximates $\dot{q}(t)$, he can infer *p(t)* (or the sequence *{p(Δt/2), p(3Δt/2),…, p((n-1)Δt/2)}*) using Hamilton's equation $\dot{q} = H_p$. Similarly, if *A* knows both *p(t)* and $\dot{p}(t)$, he can infer *q(t)* using $\dot{p} = -H_q$. I note that this is a relation between *possible* paths; *A* and *B* may still observe actual paths that are different when compared in a common representation. Because both *A* and *B* could measure or infer *q(t)*, if they measured their respective *p(t)* and *q(t)*, the probability for a given *q(t)* is *β[q]β[q]= β²[q]*. Similarly, the probability for a given *p(t)*, measured by *A* and *B*, is *α[p]α[p]= α²[p]*. The probability *Pr* for a particular path between specified points at *t$_i$* and *t$_f$*, e.g., *(q$_i$, t$_i$)* and *(q$_f$, t$_f$)*, is the probability that *A and B* observe the path in terms of *q(t) or* that *A and B* observe the path in terms of *p(t)*, i.e., *Pr = β[q]β[q] + α[p]α[p]*.

The above result shows that if *β[q]* were negative, the probability would still be positive, and an interpretation of *B's* distribution *β[q]* (or *A's* distribution *α[p]*) as probabilities is unnecessarily restrictive. I therefore generalize the notion of the distribution, calling it a mixed path to include negative numbers so that each path's *p(t)* and *q(t)* representation has a real number associated with it: *-1≤β≤1* and *-1≤α≤1*. *α²* (or *β²*) is the probability for *A* and *B* to observe a given *p(t)* (or *q(t)*). Below I will interpret the numbers *β[q]* and *α[p]* as *B's* or *A's* affinity for a path, i.e., *A* may be attracted to a "good" path with a positive value of α or repelled from a "bad" path.

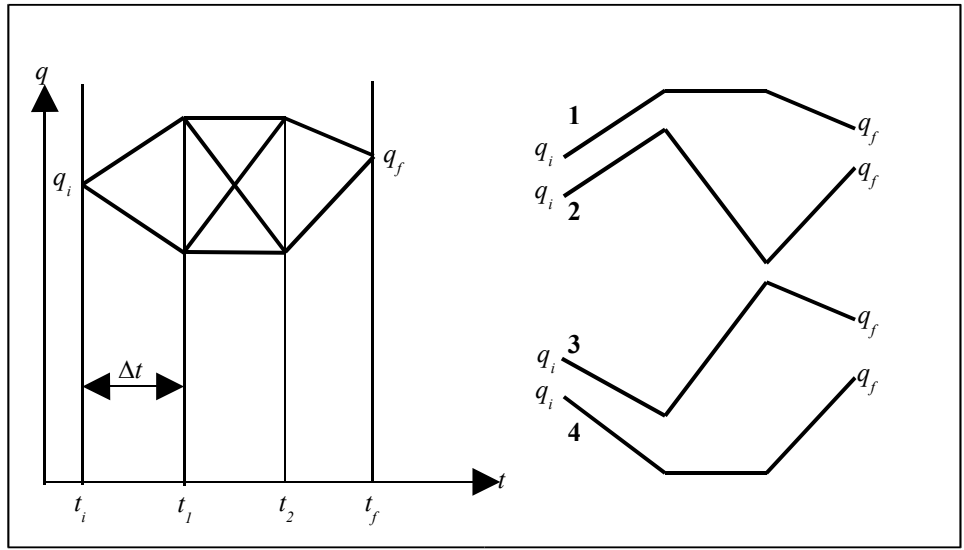

Figure 3. Six spatial-temporal points have four possible paths labeled **1**, **2**, **3**, and **4**, as shown above. The probability to observe any of the paths in terms of *q* is $\Sigma_k β[q_k] \Sigma_j β[q_j] = \{\Sigma_j β[q_j]\}^2$. Because the paths could also be observed in terms of *p*, the total probability to observe any of the four possible paths is *Pr= $\{\Sigma_j α[p_j]\}^2 + \{\Sigma_j β[q_j]\}^2$*.

I now consider two intermediate temporal points, and each of these temporal points has two possible spatial points. The four possible paths labeled **1**, **2**, **3**, and **4** are shown in Figure 3. As in the above cases, a possible path for $A$ $p(t)$ must also exist in terms of $q(t)$. Let $B$ measure a given path, say path **1**, that I call $q_1$. $A$ could measure any of the possible paths $q_1$, $q_2$, $q_3$, or $q_4$. The probability of $B$ observing $q_1$ and $A$ observing any of the paths in terms of $q$ is $\beta[q_1]\beta[q_1]+\beta[q_1]\beta[q_2]+\beta[q_1]\beta[q_3]+\beta[q_1]\beta[q_4] = \beta[q_1]\sum_j\beta[q_j]$ for $j=1,2,3,4$. The probability to observe any of the paths in terms of $q$ is $\sum_k\beta[q_k]\sum_j\beta[q_j] = \{\sum_j\beta[q_j]\}^2$. Similarly, the probability to observe any of the four paths in terms of $p$ is $\sum_k\alpha[p_k]\sum_j\alpha[p_j] = \{\sum_j\alpha[p_j]\}^2$, for $j=1,2,3,4$. The total probability $Pr$ to observe any of the four possible paths between the given beginning and ending spatial points is $Pr(q_i,q_f)=\{\sum_j\alpha[p_j]\}^2+\{\sum_j\beta[q_j]\}^2$. This can also be written in terms of the relative probability with respect to $Pr$, i.e., given that these are the only possible endpoints $\{\sum_j\alpha[p_j]\}^2+\{\sum_j\beta[q_j]\}^2=1$.

I can now allow a fan out between paths at any instant and any position in the interval $T$ by letting $\Delta t \to 0$ and $n \to \infty$ to define a path integral $\sum_j\alpha[p_j] \to \int Dp\alpha[p(t)]$ and $\sum_j\beta[q_j] \to \int Dq\beta[q(t)]$, where $Dq$ and $Dp$ are appropriately defined measures of the integral. I can find the probability $Pr$ for a particle to start at $q_i$ and end at $q_f$ by summing over all possible paths or performing the path integrals $\int Dq(t)$ and $\int Dp(t)$ to get $Pr(q_i,q_f) = \{\int Dp\alpha[p(t)]\}^2 + \{\int Dq\beta[q(t)]\}^2$.

**Generalizing Hamilton's principle**

There are two ways to observe a path through a specific $p$ or $q$. Experimenters $A$ and $B$ know that during a small time interval each possible path will extremize the functionals $R[p,q]$ and $S[p,q]$ respectively. When the initial and final points in the time intervals are not completely known there are many possible paths; neither $A$ nor $B$ can generally extremize $R[p,q]$ and $S[p,q]$ through a selection of a single "best" $p(t)$ or $q(t)$. I now show the generalization of Hamilton's principle that governs the form of the "best" pair of mixed paths $\alpha[p]$ and $\beta[q]$. Here as in the above I simplify the analysis by choosing a discrete set of points at given times, as shown in Figure 4 for three paths with fixed $q_i$ at $t_i$ and $q_f$ at $t_f$. Because any path, indexed by $j$ for $A$ and $k$ for $B$, can be represented in terms of $p(t)$ or $q(t)$, there is the same number $n$ of possible $p$ paths and $q$ paths in this subset. Corresponding to each $p_j(t)$ or $q_k(t)$ is $\alpha[p_j]=\alpha_j$ and $\beta[q_k]=\beta_k$. I represent these sets with the vectors $\alpha=(\alpha_1, \alpha_2, ..., \alpha_n)$ and $\beta=(\beta_1, \beta_2, ..., \beta_n)$. I can also define a square matrix for the action $S_{jk}$ corresponding to all possible actions for the $n$ paths with given $q_i$ and $q_f$. The elements of this action matrix $S_{jk}$ are real values of the action evaluated at paths $p_j$ and $q_k$, i.e., $S[p_j, q_k] = S_{jk}$. In the three path case, shown in Figure 4, the action matrix $\underline{S}$ is written

$$\underline{S} = \begin{Vmatrix} S_{11} & S_{12} & S_{13} \\ S_{21} & S_{22} & S_{23} \\ S_{31} & S_{32} & S_{33} \end{Vmatrix}.$$

The mixed paths are three component vectors: $\alpha=(\alpha_1, \alpha_2, \alpha_3)$ and $\beta=(\beta_1, \beta_2, \beta_3)$. I combine $\alpha$, $\beta$, and $S_{jk}$ to form the generalized action $\alpha^T \underline{S} \beta$ where $\alpha^T$ is the transpose of $\alpha$. In analogy to the case with perfect information, I extremize this generalized action by finding optimal mixed paths $\alpha_0$ and $\beta_0$. $A$ selects $\alpha_0$ so as to extremize $\alpha_0^T \underline{S} \beta$ for any given $\beta$, and $B$ selects $\beta_0$ to extremize $\alpha^T \underline{S} \beta_0$ for any given $\alpha$. When optimal mixed paths $\alpha_0$ and $\beta_0$ are found, the generalized action is $\alpha_0^T \underline{S} \beta_0 = \nu$. Because $\alpha_j$ and $\beta_k$ may be negative or positive, $\nu$ may not be the expectation value of $\underline{S}$. If $p_i$ and $p_f$ are given, I can similarly define both $\alpha[p]$ and $\beta[q]$, and the two optimal distributions $\alpha_0$ and $\beta_0$ will extremize $\alpha^T \underline{R} \beta$.

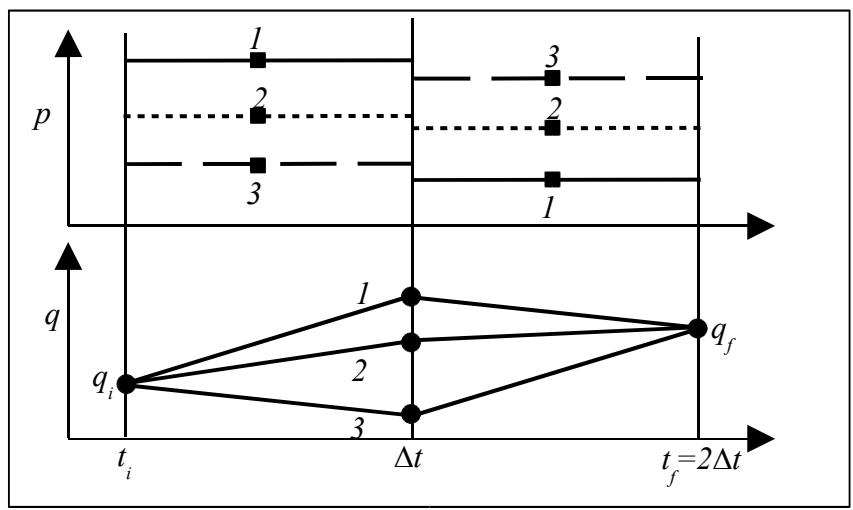

Figure 4. Five spatial-temporal points allow three possible paths. The three $p_k(t)$ paths and the corresponding three $q_j(t)$ paths have distribution vectors $\alpha=(\alpha_1, \alpha_2, \alpha_3)$, $\beta=(\beta_1, \beta_2, \beta_3)$, and an action matrix $S[p_j, q_k] = S_{jk}$. $A$ and $B$ attempt to extremize the generalized action $\alpha^T \underline{S} \beta$ by selecting optimal distributions $\alpha_0$ and $\beta_0$ such that the generalized action is extremized.

I now show there is a stationary result for a system with $n$ possible paths. I first note that the normalization condition introduces one constraint. I can account for it by letting the *2n* dimensional real vector $(\alpha, \beta)$ terminates on a *2n-1* dimensional plane. When I take $Pr=1$, the normalization condition gives $(\Sigma_j \alpha_j)^2 = a^2$ and $(\Sigma_j \beta_j)^2 = b^2 = 1-a^2$ and I may write $\alpha_0^\dagger \alpha_0 + XT = a^2$, where $XT$ are the cross terms. An equal footing between $p$ and $q$ implies that $b^2 = a^2 = \frac{1}{2}$, or more generally $b^2 = a^2 = Pr/2$, i.e., a given set of paths should have the same probability to be observed independently of whether they are observed in terms of $p$ or $q$. This gives two constraints: one on $\alpha$ and one on $\beta$. I can then let the $n$ dimensional $\alpha$ vector terminate on an *n-1* dimensional plane. The $n \times n$ real matrix $\underline{S}$ operates on the $\beta$ vector resulting in another vector in the $n$ space, as shown in Figure 5 for a 2-D vector space. The optimal mixed paths satisfy $\alpha_0^\dagger \underline{S} \beta_0 = \nu$, so I may write $\alpha_0^\dagger \underline{S} \beta_0 = \nu (\alpha_0^\dagger \alpha_0)/(a^2 - XT)$. This last expression is equivalent to $\alpha_0^\dagger [\underline{S} \beta_0 - \nu \alpha_0/(1 - b^2 - XT)] = 0$. Since by assumption $\alpha_0 \neq 0$, either $\alpha_0$ is orthogonal to $[\underline{S} \beta_0 - \nu \alpha_0/(1-b^2-XT)]$ or $[\underline{S} \beta_0 - \nu \alpha_0/(1-b^2-XT)] = 0$. Choosing the later case I get $\underline{S} \beta_0 = \nu \alpha_0/(1-b^2-XT)$ showing that the vectors $\underline{S} \beta_0$ and $\alpha_0$ are parallel (or anti-parallel) when $\upsilon$ is positive (or negative). The dashed line in Figure 5 illustrates half of the possible values of $\alpha$ for two possible paths (the other half are on an inverted line through the negative quadrant). If $\alpha_0$ and $\beta_0$ are optimal, then $\delta v=0=\delta\alpha_0^\dagger \underline{S}\beta_0+\alpha_0^\dagger \underline{S}\delta\beta_0$. Because $\alpha_0$ and $\underline{S}\beta_0$ are parallel (or anti-parallel), the first term is zero if $\delta\alpha_0$ is orthogonal to $\alpha_0$. This is so if all the components of $\alpha_0$ are equal, as shown in Figure 5. Similarly, if all the components of $\beta_0$ are equal, then $\delta\beta_0$ is perpendicular to $\beta_0$, so that the second term is also equal to zero. Because both $\delta\alpha_0^\dagger \underline{S}\beta_0$ and $\alpha_0^\dagger \underline{S}\delta\beta_0$ are separately zero, equivalence between $p$ and $q$ is maintained with this pair of optimal vectors. This solution is analogous to the fundamental hypothesis in equilibrium statistical mechanics because the affinities for each possible path are equal.

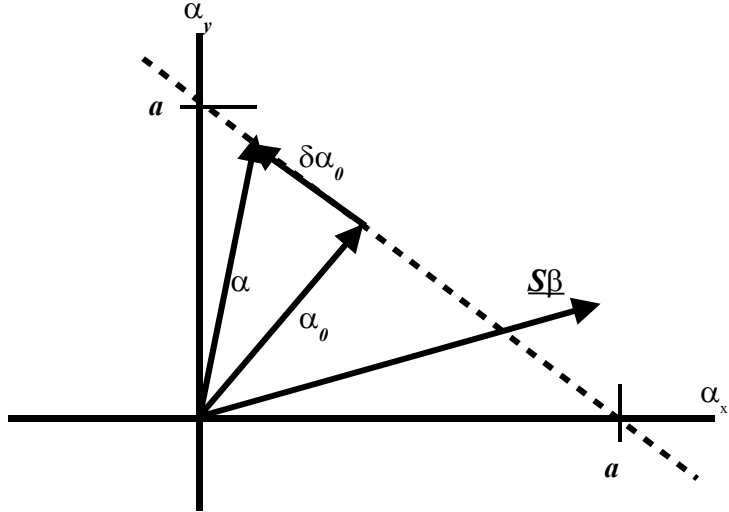

Figure 5. Shown are the vectors $\alpha$, $\underline{S}\beta$, and an optimal vector $\alpha_0$. Half of the possible values of $\alpha$ must terminate at the dashed line and the other half must terminate on the inverted line through the negative quadrant. Each component of this optimal $\alpha_0$ (or $\beta_0$) is equal.

**Game analogy and the mini-max theorem**

I next consider another extrema of the real valued action $P(\alpha,\beta)=\alpha^\dagger \underline{S}\beta$ using the following theorem from reference [6]:
Let $X$ and $Y$ be compact subsets of topological spaces, and let $P$ be a continuous function: $P:X\times Y\to R$. Then $\max_{x\in X}\{\min_{y\in Y} P(x,y)\} = \min_{x\in X}\{\max_{y\in Y} P(x,y)\} \Leftrightarrow$ there exists an $x_0$, $y_0$ and $v \in R$ such that $P(x_0,y)\geq v\ \forall\ y\in Y$ and $P(x, y_0)\leq v\ \forall\ x\in X$.

I may satisfy the compactness condition by restricting the domain of $\alpha$ and $\beta$ to be only the cases where all components have the same sign, i.e., $\alpha$ and $\beta$ are only in the upper right and lower left quadrants of Figure 5. I may include cases where $\alpha_0$ and $\underline{S}\beta_0$ are not parallel and write $\delta v=0=\delta\alpha_0^\dagger \underline{S}\beta_0 + \alpha_0^\dagger \underline{S}\delta\beta_0$ so that $\delta\alpha_0^\dagger \underline{S}\beta_0=-\alpha_0^\dagger \underline{S}\delta\beta_0$. I write $\alpha^\dagger=\alpha_0^\dagger+\delta\alpha_0^\dagger$ and $\beta=\beta_0+\delta\beta_0$. Then:
$P(\alpha_0,\beta)=\alpha_0^\dagger \underline{S}\beta=\alpha_0^\dagger \underline{S}(\beta_0+\delta\beta_0)=v+\alpha_0^\dagger \underline{S}\delta\beta_0=v-\delta\alpha_0^\dagger \underline{S}\beta_0$ and
$P(\alpha,\beta_0)= \alpha^\dagger \underline{S}\beta_0= (\alpha_0^\dagger +\delta\alpha_0^\dagger)\underline{S}\beta_0=v+\delta\alpha_0^\dagger \underline{S}\beta_0$.

This shows that the necessary and sufficient condition for the above extremum is met if $\delta\alpha_0^\dagger \underline{S}\beta_0$ always has a definite sign (this includes both the mini-max and the maxi-min cases). However one can easily see, e.g., from Figure 5, that when $\alpha_0$ and $\underline{S}\beta_0$ are not parallel $\delta\alpha_0^\dagger \underline{S}\beta_0$ can have either sign depending on the sense of the change from optimal $\delta\alpha_0$. There is, however, an additional freedom in choosing vectors because for every vector of the type $\alpha=(\alpha_x, \alpha_y)$ or $\underline{S}\beta=(\underline{S}\beta_x, \underline{S}\beta_y)$ there is also its inversion $\alpha'=(-\alpha_x, -\alpha_y)$ or $\underline{S}\beta'=(-\underline{S}\beta_x, -\underline{S}\beta_y)$. If I let $\alpha_0$ and $\beta_0$ be positive when $\delta\alpha_0$ is directed to the upper left from $\alpha_0$ and I let $\alpha_0$ and $\beta_0$ be negative when $\delta\alpha_0$ is directed to the lower right from $\alpha_0$, then I can maintain a definite sign for $\delta\alpha_0^\dagger \underline{S}\beta_0$. $P$ is continuous under inversion of $\alpha_0$ and $\beta_0$ because I let $\delta\alpha_0^\dagger$ go to zero just when the inversion is done and the theorem is satisfied.

As the reader my have noticed, the theorem cited above that led to this mini-max extremum came from Game Theory [6], [7]. This theory has recently been generalized into the complex domain in order to understand problems in quantum computation and communication [8], [9], [10], [11], [12]. I now discuss some analogous features. First of all, the players in a competitive game choose strategies that specify how a player makes his moves. These strategies are analogous to the paths of the experimenter's variable. The competing players choose strategies to maximize their payoff in analogy to the experimenter's paths extremizing the action. $S[p,q]$ is analogous to the payoff in a zero sum game where the two competing "players" have "strategies" and after each player chooses a strategy there results in a particular "payoff" to each player. The game is called zero-sum because a gain in payoff to one of the players is a loss in payoff to the other player (the sum of the two players payoff is zero). In our analogous competition each "player" chooses a path $p(t)$ or $q(t)$, respectively, and the "payoff" is the action. In fact, a particular solution to many types of games is called a "saddle point". This particular pair of strategies that each competitor chooses guaranties a certain payoff to himself (player *A*) regardless of the choice that the other player (player *B*) makes. Player *A* (who wants to maximize his payoff) picks a strategy that places the largest lower bound on his payoff. In this way, he has a guarantee of at least the amount of the largest lower bound regardless of the other player's choice. Player *B* also chooses a strategy that guarantees the smallest upper bound in *A*'s payoff (equivalent to the largest lower bound on his own payoff) and *A* can not get more payoff regardless of *A*'s choice. If the smallest upper bound and the largest lower bound are equal, then the game is said to have a saddle point solution. This mini-max character of the payoff is analogous to the extremum of the action *S*. In fact, it has been proven that there exists a saddle point solution for all finite, zero sum, two player games, with perfect information [6], [7]. Finite refers to the number of possible strategies and perfect information means that each player always knows the other player's moves during a game, e.g., tick-tack-toe, chess, etc. This is analogous to a system in classical mechanics that has perfectly known initial conditions and a unique path in phase space.

Without perfect information, the smallest upper bound and the largest lower bound in a game may not be equal. In this case, the two player zero sum game always has a saddle point in expectation. This saddle point in expectation is guaranteed for certain types of finite zero sum games (e.g., parlor games) by the Mini-max theorem, the fundamental theorem in the theory of games. This generalization to expectation values requires a definition of a mixed strategy as a probability distribution over the possible strategies. There is a mixed strategy for player *A* that guaranties the largest *lower* bound in payoff expectation to *A* independent of the choice of *B*. Conversely, there is a mixed strategy for player *B* that guarantees the smallest *upper* bound in payoff expectation to player *A* regardless of the choice of *A*. This allows a natural generalization of the saddle point between pure strategies to a saddle point between mixed strategies. I have presented an analogous generalization in the above. The mixed strategy is analogous to the mixed path.

In zero sum games the payoff to player *A* is the negative of the payoff to player *B*, and the two players have exactly opposing interests. In our analogy, I can identify the "payoff" to the "players" by noting the way each experimenter extremizes their action. As shown in reference [5], when *H(p,q)* is a saddle function the "player" choosing "strategy" *q(t)* (experimenter *B*) wants to minimize *S[p,q]* through a choice of *q(t)* subject to $\delta q(t_i) = \delta q(t_f) = 0$. The "player" choosing "strategy" *p(t)* (experimenter *A*) wants to maximize *R[p,q]* and he chooses *p* subject to $\delta p(t_i) = \delta p(t_f) = 0$. Because *S[p,q]* and *R[p,q]* are related through partial integration, they can be put into the zero sum form:

$$S = \int_{t_i}^{t_f} \left( p\dot{q} - H(p,q) \right) dt = \int_{t_i}^{t_f} \left( \frac{d[pq]}{dt} - q\dot{p} - H(p,q) \right) dt$$
$$= R + [pq]_{t_f} - [pq]_{t_i}.$$

I define the new quantities *S′* and *R′* :
$$S' = -S; \quad R' = R + [pq]_{t_f} - [pq]_{t_i}; \quad S' + R' = 0.$$

For the saddle Hamiltonian *B* determines *q* to maximize *S′* and *A* determines *p* to minimize *S′* subject to $\delta q(t_i) = \delta q(t_f) = 0$. In the above, *R′* is the "payoff" to "player" *A* and *S′* is the "payoff" to "player" *B*. When the beginning and ending conditions are exactly known by both *A* and *B*, then both *δS'=0* and *δR'=0*. This appears to be a good analogy to the saddle point solution to a zero sum game with perfect information. It is then natural to generalize this case to mixed paths with a generalized action and find the saddle point in generalized action in parallel with the Mini-max theorem. It is important, however, to emphasize that the normalization condition is quite different. In the theory of games a mixed strategy is represented by a probability distribution over strategies. If player *A* has *n* strategy choices and player *B* has *m* strategy choices, then this normalization condition corresponds to $\Sigma_j \alpha_j = 1$ and $\Sigma_k \beta_k = 1$, where α and β are the probabilities of the strategies and *j=1,2,…,n; k=1,2,…,m* are the indices for the strategy choice. Our normalization condition is a single condition that does not require positive values for the elements of the mixed paths.

This special choice of optimal vectors in the restricted domain, discussed above, produces a mini-max (or maxi-min) extremum. In an unrestricted domain there may exist solutions where individual elements of an optimal vector may be negative or positive. The element, e.g., $S_{12}$ may be large and $S_{12}\beta_2$ would seem to be a large potential payoff to *B*. *A*'s element $\alpha_1$, however, may be negative to create a potentially large loss to *B* $\alpha_1 S_{12} \beta_2$. This concept of the mixed path allows *A* or *B* to select a negative value for an element of their mixed path in order to "negate" a large expected value of the action for the other experimenter. Above I interpreted the numbers *α[p]* and *β[q]* as a system's affinity for a path. *A* or *B* may be attracted or repelled from a path because of the large expected value of the generalized action that the other "player" may get and may negate a large expected value to the other player.

**Probability amplitude**

As an example, consider a system with two possible paths with affinities $\alpha_0 = (\alpha_1, \alpha_2)$ for paths $(p_1(t), p_2(t))$ and $\beta_0 = (\beta_1, \beta_2)$ for $(q_1(t), q_2(t))$. I first note that the above optimal vectors give $\alpha_0^\dagger \underline{S} \beta_0 = \alpha_1 S_{11} \beta_1 + \alpha_1 S_{12} \beta_2 + \alpha_2 S_{21} \beta_1 + \alpha_2 S_{22} \beta_2 = (S_{11} + S_{12} + S_{21} + S_{22})\alpha_1 \beta_1 = \nu$ or $s = \nu/(\alpha_1 \beta_1)$, where $s = S_{11}+S_{12}+S_{21}+S_{22}$. I can combine the two distributions for path 1, $\alpha_1$ and $\beta_1$, into a complex quantity $\varphi_1 = \alpha_1 + i\beta_1$ and similarly for path 2 $\varphi_2 = \alpha_2 + i\beta_2$. The normalization condition is written as $(\Sigma_j \alpha_j)^2 + (\Sigma_k \beta_k)^2 = |(\Sigma_j \phi_j)|^2 = 1$. This complex representation consolidates the two real vectors $\alpha_0$ and $\beta_0$ into one complex vector $\phi$, where each component is the probability amplitude for a particular path. Next I consider a functional for a physical quantity of interest defined with an initial and final point $X[p,q]$ (e.g., the action $S[p,q]$ or the energy $E[p,q] = \int H(p,q)dt$) that reduces to a 2X2 matrix with two possible paths. I can form the quantity $<X> = \phi^{*\dagger}\underline{X}\phi = \phi_1^* X_{11} \phi_1 + \phi_1^* X_{12} \phi_2 + \phi_2^* X_{21} \phi_1 + \phi_2^* X_{22} \phi_2 = \alpha_1 X_{11} \alpha_1 + \alpha_2 X_{22} \alpha_2 + \beta_1 X_{11} \beta_1 + \beta_2 X_{22} \beta_2 + \phi_1^* X_{12} \phi_2 + \phi_2^* X_{21} \phi_1$. The imaginary part of these last two cross terms are zero because $\beta_2 = \beta_1$ and $\alpha_1 = \alpha_2$ for the optimal vectors discussed above, i.e., $Im[\phi_1^* X_{12} \phi_2 + \phi_2^* X_{21} \phi_1] = (X_{12} - X_{21})(\alpha_1 \beta_2 - \alpha_2 \beta_1) = 0$. The final result is $<X> = \alpha_1 X_{11} \alpha_1 + \alpha_2 X_{22} \alpha_2 + \beta_1 X_{11} \beta_1 + \beta_2 X_{22} \beta_2 + (\alpha_1 \alpha_2 + \beta_2 \beta_1)(X_{21} + X_{12})$. This real quantity may be interpreted as the expectation value of $X[p,q]$, where the first four terms express the expectation that $A$ and $B$ observes path *1* or *2* weighted by the associated $X_{ii}$. More interestingly, this expression gives weight to the possibility that paths *1* and *2* in terms of $p$ may be observed simultaneously by $A$ and $B$, or that paths *1* and *2* in terms of $q$ may be seen simultaneously. I may interpret these last four terms to be the quantum mechanical interference terms that could increase or decrease $<X>$ depending on the values of the $\alpha$'s and $\beta$'s. These combinations are possible because $p$ and $q$ are not known simultaneously. With the help of the game analogy, the interference is interpreted as one of the players decreasing or increasing the expected value through the selection of his components. In $N$ dimensions the expectation is $<X> = \sum_{i,j=1}^{N} (\alpha_i X_{ij} \alpha_j + \beta_i X_{ij} \beta_j)$.

The basis that contains the optimal vectors $\alpha_0$ and $\beta_0$ are not necessarily the same. Even though $\alpha_0$ is parallel to $\underline{S}\beta_0$, $\alpha_0$ is parallel to $\beta_0$ only in the extraordinary case when $\underline{S}$ is diagonal. I may combine these two vectors using common parameters if they are expressed in a common basis. Any basis where $\alpha_0$ and $\beta_0$ are represented may then be specified, in the 2-D case, by two angles $\theta_1$ and $\theta_1'$, as shown in Figure 6, where $\theta_1$ and $\theta_1'$ specify the angle between $\alpha_0$ and $\beta_0$ and the $x_1$ - axis. A convenient representation is a basis where $\alpha_0$ and $\beta_0$ are symmetric about the $\pi/4$ line, as shown in Figure 6. In this case, $\{\alpha_1, \beta_1\}$ is $\{|a|cos(\theta_1), |a|sin(\theta_1)\}$ and $\{\alpha_2, \beta_2\}$ is $\{|a|cos(\theta_2), |a|sin(\theta_2)\}$, where $\theta_1 + \theta_1' = \pi/2$ and $\theta_2 + \theta_2' = \pi/2$. The angles $\theta_2$ and $\theta_2'$ specify the angle between $\alpha_0$ and $\beta_0$ and the $x_2$-axis. The complex quantities for path 1 and path 2 are $\phi_1 = \alpha_1 + i\beta_1 = |a|cos(\theta_1) + i|a|sin(\theta_1) = |a|exp(i\theta_1)$ and $\phi_2 = \alpha_2 + i\beta_2 = |a|cos(\theta_2) + i|a|sin(\theta_2) = |a|exp(i\theta_2)$. $\phi_1$ and $\phi_2$ are the probability amplitudes for path *1* and *2* respectively. This same analysis is easily generalized to many paths, e.g., for three paths the two vectors with a given angle between them can be oriented with 3 angles in a 3 dimensional vector space that is fixed by three constraints between the corresponding angles (or direction cosines), $\theta_i + \theta_i' = \pi/2$, $i=1,2,3$. In $N$ dimensions there are $N$ degrees of freedom and $N$ constraints between corresponding angles. I can then find the total probability amplitude $K(q_i,t_i;q_f,t_f)$ as the sum over all $\phi_j$, i.e., $K(q_i,t_i;q_f,t_f) = \Sigma\phi_j = a\Sigma exp(i\theta_j)$ between the given endpoints. In the quantum mechanical domain $\theta_j=(2\pi/h)S_j$, where $h$ is Planck's constant, and the propagator is $K(q_i,t_i;q_f,t_f)$ [13]. The usual argument for the additivity of $S_j$ between consecutive endpoints implies the multiplying of $K$'s between consecutive endpoints. This in turn leads to the Schrödinger Wave equation and is the essential ingredient for the path integral formulation of quantum mechanics [13] and quantum field theory [14]. In this case, the Heisenberg uncertainty principle $h \geq \delta p \delta q$ gives us the measure of the uncertainty for phase space points and tells us the lower limit of imperfect information. This formulation, however, is more general and should also be applicable in cases when $\delta p \delta q >> h$.

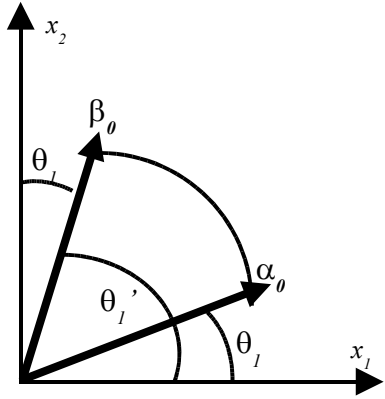

Figure 6. By combining the two optimal vectors $\alpha_0$ and $\beta_0$ in a common basis using two angles $\theta_1$ and $\theta_1'$, where $\theta_1 + \theta_1' = \pi/2$ are the angles from $\alpha_0$ and $\beta_0$ to the $x_1$ axis and $\theta_2 + \theta_2' = \pi/2$ are the angles to the $x_2$ axis, I can then construct a complex quantity for path 1 and path 2, i.e., $\phi_1 = \alpha_1 + i\beta_1 = |a|\cos(\theta_1) + i|a|\sin(\theta_1) = |a|\exp(i\theta_1)$ and $\phi_2 = \alpha_2 + i\beta_2 = |a|\cos(\theta_2) + i|a|\sin(\theta_2) = |a|\exp(i\theta_2)$. These complex numbers $\phi_1$ and $\phi_2$ are the probability amplitudes for path $1$ and $2$ respectively. I can find the propagator $K(q_i,t_i;q_f,t_f)$ as the sum over all $\phi_j$ $K(q_i,t_i;q_f,t_f) = \Sigma_j\phi_j = a\Sigma exp\{(2\pi/h)S_j\}$ between the given endpoints.

**Quantum mechanical propagator**

In a system where there is not perfect simultaneous knowledge of $p$ and $q$, I have hypothesized that Hamilton's principle is still a general principle of physics. As discussed above, its application to systems with imperfect information can be accomplished through an extremum where the distribution functional for $p(t)$ and $q(t)$ satisfy an extremum principle that is analogous to a mini-max extremum of a zero-sum game. The optimal distributions (mixed paths) satisfy an extremum condition for $\alpha^t S\beta$, the generalized action for $B$, subject to the normalization condition for the mixed paths $\alpha[p]$ and $\beta[q]$. A direct generalization of the previous discussion gives Hamilton's principle as an extremization over distribution functionals of a path integral

$\underset{\alpha[q]}{Ext}\left(\underset{\beta[p]}{Ext}\iint_{p,q} DqDp\alpha[p] \, S[p,q] \, \beta[q]\right) = v$, where $\underset{X[y]}{Ext}$ represents the extremum over the functional $X[y]$. The distributions that satisfy the above condition ($\alpha_0[p]$, $\beta_0[q]$) are the actual distribution functionals. Such a pair of actual distributions must also satisfy the equivalent pair of conditions:

$$\underset{\beta[q]}{Ext}\left(\int_q Dq\beta[q]\int_p Dp\alpha_0[p]\, S[p,q]\right) = \nu, \text{ and}$$

$$\underset{\alpha[p]}{Ext}\left(\int_p Dp\alpha[p]\int_q Dq\beta_0[q]\, S[p,q]\right) = \nu. \quad \text{Defining} \quad S_\alpha[q] = \int_p Dp\alpha_0[p]S[p,q] \quad \text{and}$$

$$S_\beta[p] = \int_p Dq\beta_0[q]S[p,q], \text{ these conditions can be rewritten as}$$

$$\underset{\beta[q]}{Ext}\left(\int_q Dq\beta[q]S_\alpha[q]\right) = \underset{\alpha[p]}{Ext}\left(\int_p Dp\alpha[p]S_\beta[p]\right) = \nu. \qquad (1)$$

I can guess a solution ($\alpha_0[p]$, $\beta_0[q]$) and test if it satisfies equation (1). In particular, I may take each possible path to have equal affinity (to be "equally likely") as suggested in the above finite dimensional case. Expressing $\alpha$ and $\beta$ in a common basis, however, still requires direction cosine variables that may be associated with the action. I found in reference [5] that the functional $J[q(t)]$ could be found from $S[p(t),q(t)]$ by eliminating $p(t)$ from $S$ using the solution to one of Hamilton's equations $\dot{q} = H_p$. Another functional $G[p]$ is obtained from $S$ using the other Hamilton's equation $\dot{p} = -H_q$ to eliminate $q(t)$. Based on our previous discussion it is natural to use these functionals to separately map the paths $p(t)$ and $q(t)$ to real numbers. The distribution functionals can then be written as functions of these functionals $\alpha_0(G[p])$ and $\beta_0(J[q])$), i.e., I assume that all paths with the same $J$ or $G$ have the same affinity. The distribution functionals must also satisfy the normalization condition that can be expressed as $\{\int Dp\alpha[p(t)]\}^2 + \{\int Dq\beta[q(t)]\}^2 = Pr$. One choice of $\alpha$ and $\beta$ that may satisfy the above conditions is $\alpha = a\sin(cG[p])$ and $\beta = b\cos(cJ[q])$, where $a$, $b$, and $c$ are constants to be determined. This choice also generalizes the discussion above for $N$ paths to an infinite number of paths. Although the two amplitudes $|a|$ and $|b|$ must be equal for each path to have equal affinity, I include them here because there is still a sign ambiguity in the affinities $\alpha$ and $\beta$. I will now treat $\alpha$ and $\beta$ as functions of the parameters $(a,b,c)$ and determine them with the help of Lagrangian multipliers. I have two problems.

**Problem 1)** is to find $\underset{\alpha[p]}{Ext}\left\{\int_p Dp\alpha[p]S_\beta[p]\right\} = Ext_{(a,c)}\{\int Dpa\sin(cG[p])S_\beta[p]\}$. This is subject to the constraint $\{\int Dp\alpha[p]\}^2 + \{\int Dq\beta_0[q]\}^2 = Pr$ or in terms of relative probability $\{\int Dp\alpha[p]\}^2 + \{\int Dq\beta_0[q]\}^2 = 1$. I want to extremize $f_1(a,c) = a\int Dp\sin(cG[p])S_\beta[p] = aI_1(c)$. This extremization is subject to the constraint

$$g_1(a,c) = a^2\{\int Dp\sin(cG[p])\}^2 + \{\int Dq\beta_0[q]\}^2 - 1 = a^2 J_1^2(c) - \kappa_1^2 = 0.$$

This problem is easily done with three equations and three unknowns. I define $F_1(a,c) = f_1 + \lambda_1 g_1$. The critical $(a^*, c^*, \lambda_1)$ are found using the usual conditions, i.e., $\left[\frac{\partial F_1}{\partial a}\right]_{a^*} = 0$, $\left[\frac{\partial F_1}{\partial \lambda_1}\right]_{\lambda_1} = 0$, and $\left[\frac{\partial F_1}{\partial c}\right]_{c^*} = 0$. The first two equations yield $a^* = \pm\frac{\kappa_1}{J_1}$ and $\lambda_1 = \dfrac{\mp I_1}{2J_1\kappa_1}$. The final equation is $\left[\dfrac{\partial F_1}{\partial c}\right]_{c^*} = \left[a\dfrac{\partial I_1}{\partial c} + a^2\lambda_1 2J_1\dfrac{\partial J_1}{\partial c}\right]_{c^*} = 0$. Substituting the above results gives the differential equation

$\left[\dfrac{1}{I_1}\dfrac{\partial I_1}{\partial c}\right]_{c^*} = \left[\dfrac{1}{J_1}\dfrac{\partial J_1}{\partial c}\right]_{c^*} \Rightarrow \left[\dfrac{d\ln(I_1)}{dc}\right]_{c^*} = \left[\dfrac{d\ln(J_1)}{dc}\right]_{c^*}$. This latter condition is equivalent to $\dfrac{d}{dc}\{\ln(I_1) - \ln(J_1)\}_{c^*} = 0 \Rightarrow \dfrac{I_1(c^*)}{J_1(c^*)} = const. = C_1$. Evaluating this last condition to find the value of $c^*$ requires a path integral.

**Problem 2)** is to find $\underset{\beta[q]}{Ext}\left\{\int_q Dq\,\beta[q]S_\alpha[q]\right\} = Ext_{(b,c)}\{\int Dq\,b\cos(cJ[q])S_\alpha[q]\}$. This is subject to the constraint $\{\int Dp\alpha_0[p]\}^2 + \{\int Dq\beta[q]\}^2 = 1$. I want to extremize $f_2(b,c) = b\int Dq\cos(cJ[q])S_\alpha[q] = bI_2(c)$ subject to $g_2(b,c) = b^2\{\int Dq\cos(cJ[q])\}^2 + \{\int Dp\alpha_0[p]\}^2 - 1 = b^2J_2^2(c) - \kappa_2^2 = 0$.

I define $F_2(b,c) = f_2 + \lambda_2 g_2$. The critical $(b^*, c^*, \lambda_2)$ are found using the same conditions, i.e., $\left[\dfrac{\partial F_2}{\partial b}\right]_{b^*} = 0$, $\left[\dfrac{\partial F_2}{\partial \lambda_2}\right]_{\lambda_2} = 0$, and $\left[\dfrac{\partial F_2}{\partial c}\right]_{c^*} = 0$ giving $b^* = \pm\dfrac{\kappa_2}{J_2}$, $\lambda_2 = \dfrac{\mp I_2}{2J_2\kappa_2}$, and $\dfrac{I_2(c^*)}{J_2(c^*)} = const. = C_2$. The two constants $C_1$ and $C_2$ are both positive and are related through equation (1).

I now examine these extrema to see if they are maximum conditions or minimum conditions as we might expect from the game theory analogy. The minimum conditions for problem 1) are $D_1 = [F_{1aa}F_{1cc} - F_{1ac}^2]_{a^*c^*\lambda} > 0$ and $[F_{1aa}]_{a^*c^*\lambda} < 0$. The maximum conditions for problem 2) are $D_2 = [F_{2bb}F_{2cc} - F_{2bc}^2]_{a^*c^*\lambda_1} > 0$ and $[F_{2bb}]_{a^*c^*\lambda_1} > 0$. Evaluating the inequalities $D_1|_{a^*,c^*,\lambda} > 0$ and $D_2|_{b^*,c^*,\lambda} > 0$, I find that $D_1|_{a^*,c^*,\lambda} = D_2|_{b^*,c^*,\lambda} = 0$. This means that I cannot determine if either problem's extrema are a maximum or a minimum using this method. This may have been expected from the discussion above for the finite dimensional case. Nevertheless, I have shown that $\alpha_0 = a^*\sin(c^*G[p])$ and $\beta_0 = b^*\cos(c^*J[q])$ satisfy an extremum condition. The complex amplitude for a path is $\phi_0 = \alpha_0 + i\beta_0 = a^*\sin(c^*G[p]) + ib^*\cos(c^*J[q]) = a^*(\sin(c^*S[p,q(p)]) + i\cos(c^*S[p(q),q])$, where I have set $a^* = b^*$. Changing the variable back to $p$ and $q$, and the fact that each $p$ has a corresponding $q$, shows that $\phi_0 = a^*\exp(ic^*S[p,q])$. The phase angle in the discrete probability amplitude is proportional to the action $S$, as we guessed in the above section.

To actually evaluate the path integral to find $a^*$ and $c^*$, I must actually do a path integral. There are well-documented procedures for doing this and the interested reader may explore references [15], [16], and many others.

**Irreversible behavior and the micro-canonical ensemble**

Pure solutions, corresponding to simultaneous perfect knowledge of $p$ and $q$, are time reversal invariant. If at any given time the value of a perfectly known $p$ at its perfectly known position is reversed the system will retrace its path converging back to its former phase space point. Formally, Hamilton's equations have the same form when $H(p, q, t)=H(-p, q, -t)$, $p \rightarrow -p$, $q \rightarrow q$, and $t \rightarrow -t$. Consider, for example, a mole of gas in unbounded space where all the molecules interact through repulsive and conservative forces. If at some initial time the molecules are released from a small volume, they will diverge from this volume filling a larger volume with time. When the molecules are dispersed and time is reversed (or $p$ is inverted) this $\sim 10^{23}$ dimensional point in phase space will exactly reverse its path and return to the initial point. If both $p$ and $t$ are simultaneously changed, then the molecules will continue to disperse.

In a time-reversed system with imperfect information, every *possible* path will retrace its path to its *hypothetical* initial phase space point. Formally, when $p \rightarrow -p$, $q \rightarrow q$ and $t \rightarrow -t$, the $q(t)$ paths, as shown in Figure 1, are reflected about the $q$ axis and the $p(t)$ paths are inverted in $(p,t)$. The amplitude for a given path $\phi_j = exp(icS_j) = exp(i2\pi S_j/h)$ between fixed beginning and ending points will be time reversal invariant if $t \rightarrow -t$ and $i \rightarrow -i$. This latter condition corresponds to $\alpha[q] \rightarrow \alpha[-q]$ and $\beta[p] \rightarrow -\beta[-p]$. Without perfect information, time reversal invariance has all *possible p* reversed when time is reversed and the optimal mixed path is time reversal invariant if the $\alpha$ distribution is even in $q$ and the $\beta$ distribution is odd in $p$.

Between the initial and final times, however, the number of possible paths continues to increase because reversing time has not increased the available information. When there is imperfect information, the fan out of paths continues and the number of possible paths only decreases when the system arrives near the final fixed point $q_f$ or $p_f$. If this experiment were repeated with a box around the initial volume, the larger volume may be represented by a 3-D infinite-well potential and the values of the positions are constrained. The values of $p$ are also constrained by total energy $E_T$ conservation and total momentum $\mathbf{P}_T$ conservation. When the final time is extended to infinity there is never a decrease in information, but the bounds in phase space places an upper limit on the range of possible positions and momenta at a given time. The increase in the number of possible paths will continue until there is at least one possible path to each phase space point allowed by the constraints. At this time, measured by the relaxation time, I expect that a uniform distribution may be realized because all paths have equal affinity. The relaxation time characterizes the rate at which possible phase space points are created as the paths fan out. The mixed path is then time reversal invariant in the sense that each possible path is reversible and irreversible in the sense that there are more possible paths as time evolves until all the accessible phase space is filled irrespective of the direction in time.

In the space and time domain a final fixed point decreases the number of possible phase space paths, and interference effects (e.g., many paths arrives at the same point) can skew the probabilities away from uniformity. I note that the complex amplitude for a given path is $\phi_j = a exp(i\theta_j) = a exp(cS_j)$ between respective spatial-temporal beginning and ending points. The constant $c^{-1}$ characterizes the uncertainty in $p$ and $q$, e.g., in the quantum domain there is the minimum $c^{-1} = h/2\pi$ where $h$ is Planck's constant. If I take the maximum $c=2\pi/h$, then $\theta_j = cS_j = \frac{2\pi}{h}S_j$. An example is the 1-D free particle propagator $K(q_i, q_f, t_i, t_f) = \sqrt{\frac{cm}{i2\pi\Delta t}} \exp\left(\frac{icm\Delta q^2}{2\Delta t}\right)$. Because the action for $N$ free particles in 3-D is separable, the propagator is a product of *3N* of the above propagators

$$K(\vec{q}_i,\vec{q}_f,t_i,t_f) = \sqrt{\left(\frac{m}{ih\Delta t}\right)^{3N}} \exp\left(\frac{i2\pi m \Delta \vec{q}^2}{2h\Delta t}\right),$$ where $\vec{q}$ is a *3N* dimensional position

vector. The probability density *Pr* to go from $d^{3N}q_i$ about $\vec{q}_i$ at $t_i$ to $d^{3N}q_f$ about $\vec{q}_f$ at $t_f$ is $Pr(\vec{q}_i,\vec{q}_f,t_i,t_f) = KK^* = \left(\frac{m}{h\Delta t}\right)^{3N}$. *Pr* is independent of $\vec{q}_i$, $\vec{q}_f$, and constant for fixed $\Delta t$. Although this may seem to be a uniform distribution, the free particle is only a valid picture until one of the particles collides with the wall. In an infinite spatial domain $L \to \infty$ and *Pr* goes to zero as $\Delta t^{-3N}$ when $\Delta t \to \infty$. As before, I may gain some insight into the general case by using discrete paths and write

$$Pr(\vec{q}_i,\vec{q}_f,t_i,t_f) = \left(a\sum_j^n e^{i\theta_j}\right)^* \left(a\sum_k^n e^{i\theta_k}\right) = |a|^2 \sum_{j=k} 1 + |a|^2 \sum_{j \neq k} e^{i(\theta_k - \theta_j)},$$ where *n* is the

number of paths. To keep the sum finite as $n \to \infty$, the amplitude may go as $|a|^2 \sim 1/n$. The first term is then a constant and the second term oscillates with an amplitude $|a|^2$ that goes to zero, especially when there is a large number of particles as in a macroscopic system. Random superposition will destroy any contribution between paths of large $\theta_k - \theta_j$. The contribution from paths with small $\theta_k - \theta_j$ (usually paths very close to the classical path) would also tend to zero as the amplitude dampens the oscillations. This shows how a system with many degrees of freedom could give *Pr→constant*.

If I except that all that is known is $E_T$ and $\textbf{\textit{P}}_T$, then I may convert the "ideal gas" propagator $K(\vec{q}_i,\vec{q}_f,t_i,t_f) = \sqrt{\left(\frac{m}{ih\Delta t}\right)^{3N}} \exp\left(\frac{i2\pi m \Delta \vec{q}^2}{2h\Delta t}\right)$ by Fourier transforming the endpoints in $p_i$ and $p_f$ and the end times in $E_i$ and $E_f$ [13]. The result is:

$$K(\vec{p}_i,\vec{p}_f,E_i,E_f) = C\delta(\vec{p}_i - \vec{p}_f)\delta\left(E_i - \frac{\vec{p}_i^2}{2m}\right)\delta\left(E_f - \frac{\vec{p}_f^2}{2m}\right).$$

where I have used the complete ignorance in $q_i$, $q_f$, $t_i$, and $t_f$ to integrate each quantity between $\pm\infty$. This includes letting $L \to \pm\infty$, $t_i \to \pm\infty$, and $t_f \to \pm\infty$, i.e., there is complete ignorance in the direction of time. The probability density is then $Pr(\textbf{\textit{P}}_T,E_T) = KK^* = |C|^2[\delta(\textbf{\textit{p}}-\textbf{\textit{P}}_T)\delta(E-E_T)]^2$, and I arrive at the micro-canonical ensemble [4]. I note that with a large but finite volume $-\infty < L < \infty$ and a large but finite time interval the delta functions become sharply peaked and thin functions reflecting a finite fluctuation probability. These fluctuations have their origin in quantum mechanical superposition. This function approximates a delta function for macroscopic volumes and time intervals much larger than the relaxation time. The probability distribution for such an isolated system, i.e., of given volume, total energy, and total momentum, is uniform. This micro-canonical ensemble has a uniform distribution subject only to the conditions that there is a sizeable number of possible paths (or particles except in low temperature cases), that any averaging be longer than the relaxation time (the time that characterizes the fan out rate), and that the spatial domain be macroscopic (much larger than a particle).

## Conclusion

A new connection between classical mechanics and quantum mechanics is proposed. The quantum mechanical propagator is derived from a variational principle that I identify as a generalized form of Hamilton's principle. The system has imperfect information, i.e., there is incomplete knowledge of the physical state. Two distribution functionals over possible *p* paths α*[p(t)]*, over possible *q* paths β*[q(t)]*, and a generalized action corresponding to a matrix of the action evaluated at all possible *p* and *q* are defined. The generalized Hamilton's principle is the extremization over all possible distributions of $\iint_{p,q} Dq Dp \, \alpha[p] \, S[p,q] \, \beta[q]$. The normalization of the two distributions allows their values to be negative and they are identified as the real and imaginary parts of the complex amplitude. The amplitude in the Feynman path integral is then shown to contain both optimal vectors that extremize the generalized action. Applying this formulation to many particles in a macroscopic system, I show how irreversible behavior and the micro-canonical ensemble follow immediately.

**References**
[1] L. D. Landau and E. M. Lifshitz, *Mechanics* (Pergamon Press, Oxford, 1976).
[2] C. Fox, *An Introduction to the Calculus of Variation* (Dover, New York, 1987).
[3] A. M. Arthurs, *Complementary variational principles*, (Clarendon Press ; New York, 1980).
[4] L. D. Landau and E. M. Lifshitz, *Statistical Physics* (Pergamon Press, Oxford, 1980).
[5] J. Hegseth, *The path integrals from classical momentum paths*, eprint, quant-phys/0403005.
[6] A. J. Jones, *Game theory: mathematical models of conflict*, (Chichester, West Sussex, E. Horwood, 1980).
[7] M. Dresher, *The Mathematics of games of strategy: theory and applications* (Dover, New York, 1981).
[8] David A. Meyer, Phys. Rev. Lett. **82**, 1052 (1999).
[9] J. Eisert, M. Wilkens, and M. Lewenstien, Phys. Rev. Lett. **83**,3077 (1999).
[10] L. Goldenberg, L Vaidman, and S. Wiesner, Phys. Rev. Lett. **82**,3356 (1999).
[11] J. Du, et. al., Phys. Rev. Lett. **88**,137902-1 (2002).
[12] C. Lee and N. Johnson, Phys. Rev A **67**, 022311 (2003).
[13] R. P. Feynmann and A. R. Hibbs, *Quantum Mechanics and Path Integrals* (McGraw-Hill, New York, 1965).
[14] A. Zee, *Quantum Field Theory in a Nutshell* (Princeton University Press, Princeton, New Jersey, 2003).
[15] L. S. Schulman, *Techniques and Applications of Path Integration* (John Wiley and Sons, New York, 1981).
[16] H. Kleinert, *Path Integral in Quantum Mechanics, Statistics, Polymer Physics, and Financial Markets* (World Scientific, New Jersey, 2004).